\documentstyle[11pt,IAU207_pasp,twoside,psfig]{article}
\def\edcomment#1{\iffalse\marginpar{\raggedright\sl#1\/}\else\relax\fi}
\reversemarginpar

\begin{document}
\title{Extragalactic Star Clusters: Speculations on the Future}
 \author{J. S. Gallagher}
\affil{Dept.\ of Astronomy, U.\ Wisconsin, Madison, WI 53706, USA}
\author{E. K. Grebel}
\affil{Max-Planck-Institut f\"ur Astronomie,  K\"onigsstuhl 17, D-69117 
Heidelberg, Germany}

\begin{abstract}
We discuss the future possibilities for extragalactic
star cluster research with the expected new ground-based and space-based
telescopes and instrumentation.  Significant gains are expected due to
improved angular resolution, sensitivity, and area coverage 
particularly in the infrared and radio, accompanied by 
progress in evolutionary and dynamical modelling.  
Improvements in angular resolution are anticipated, especially through
new adaptive optics systems (e.g., Keck, Gemini, VLT), and interferometry 
(e.g., 
Keck, VLT, LBT, ALMA, SMA, SkA), and space instrumentation (e.g., Chandra,
NGST), enabling
studies even of deeply embedded, forming extragalactic star clusters.  
Tidal disruption of Galactic clusters
becomes observable through wide-area surveys such as
the SDSS, VISTA, PRIME, including proper motion measurements through 
high-resolution imaging (e.g., HST, LBT, SIM, GAIA).
Sensitive new optical and infrared spectrographs (e.g., HET, SALT, GranTeCan,
Magellan, Keck, VLT, CELT, OWL, NGST) will push kinematic
and abundance studies to new limits, allowing us detailed comparisons
with model predictions.
One important wavelength range for the study of young, massive star
clusters, the far UV, appears to be neglected
by future planned instrumentation.
\end{abstract}

\section{Introduction}

While research on extragalactic star clusters has accelerated during
the past decade, this field has yet to achieve full maturity. In part
this is due to the demanding angular resolution and sensitivity
requirements for measurements of star clusters in galaxies.  The
complex nature of star clusters also plays a role. Figure 1 presents a
simple analogy to the life of a star cluster. It arises from and is
shaped at birth by a variety of factors, including the structure of
the ISM and properties of its host galaxy. Once separated from its
natal gas, stellar and dynamical evolution will largely determine the
degree to which a cluster survives as a gravitationally bound
system. After several Gyr the rate of change may slow, with an
important exception in cases of cluster core collapse, and the system
should evolve passively to resemble the ancient globular star
clusters.

The evolution of a star cluster, while considerably more complicated
than that of a single star, is yet much simpler than what occurs in
galaxies. Despite their complications, star clusters are the brightest
and best samples of coeval star formation. They preserve information
about past star formation processes through their chemical abundances,
initial mass functions, and overall structures and are the mainspring
of the stellar evolution clock. Star clusters form in a range of
environments extending from extreme starbursts to the lazy birth of
stars in dwarf galaxies, e.g., Grebel (2000).  This paper briefly reviews
some aspects of the life cycles of star clusters before turning to
speculations about the future of this research area. We focus on the
early development of dense, massive clusters in our by no means
complete selection of examples. Features of globular clusters are
covered by Lee (2002), Fall (2002), and many other excellent reviews in
these proceedings.

\begin{figure}
\centerline{\vbox{
  \psfig{figure=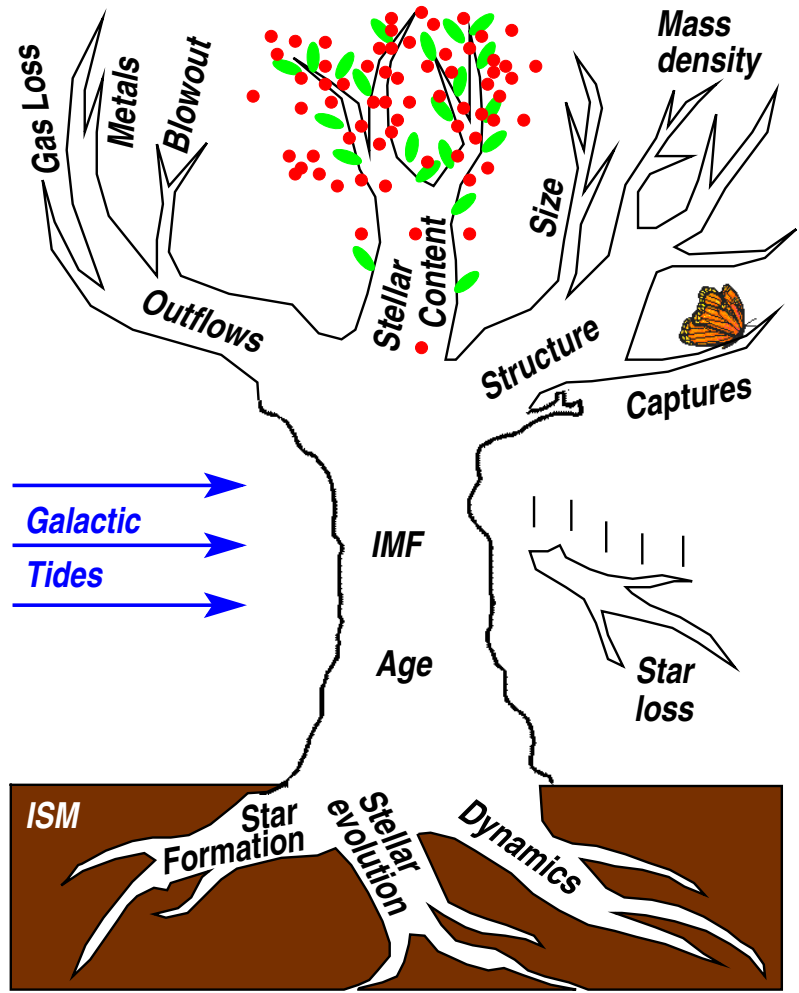,width=12cm,angle=0}
}}
\caption{A cartoon illustration of some of the factors entering into the 
observable properties of star clusters.}
\end{figure}

\section{Evolutionary Phases}

The evolution of massive star clusters has been discussed throughout 
this conference. Table 1 summarizes  
a simple model for the early evolution 
of a massive star cluster, based on the scheme outlined by 
Kobulnicky \& Johnson (private communication). Our main goal here is 
to place the evolutionary path in an observational context, so that we can 
consider how advances in technology might influence our ability to 
chart the evolution of star clusters. For this illustration we assume 
that the evolution of a region surrounding a super star cluster 
depends on the state of the cluster; e.g., the giant HII region 
appears after the ionization front from the cluster advances into the 
surrounding medium.  A key point is that 
during the first 100~Myr of the life of the cluster, it will produce 
interesting observational signatures across much of the electromagnetic 
spectrum, from radio emission during formation and as a dense, compact 
HII region, to X-rays associated with stellar winds and supernovae.

\begin{table}
\caption{Phases of Cluster Evolution}
\begin{tabular}{clrll} 
\tableline
Stage & Signature & Span & Properties & Observations \\
\tableline
{\bf I} & Pre-formation & $-$1 -- 0~Myr? & dense molecular gas & radio: mm-cm \\
{\bf II} & Stellar birth & 0 -- 1~Myr    & ultradense HII/dust & radio $+$ FIR \\
{\bf III} & Giant HII  & 1 -- 7~Myr   & HII/UV stars & UVOIR \\
{\bf IV} & Supernovae  & 3 -- 30~Myr & OB stars/SNe~II & UVOIR, X-ray \\
{\bf V} & Supergiants & $\sim$10~Myr & peak optical/UV L & UVOIR \\ 
{\bf VI} & AGB & 0.3 -- 2~Gyr & extended-AGB$+$MS & OIR $+$ FIR \\
{\bf VII} & RGB & $>$2 -- 3~ Gyr & RGB dominates & OIR \\
\tableline
\tableline
\end{tabular}
\end{table}

An equally important factor is our ability to resolve structure within
star clusters, and to distinguish them from their often complex host
galaxy backgrounds. As illustrated in Table 2, this ranges from nearly
full access to individual stars on the lower main sequence 
around the Milky Way to an
ability distinguish the most luminous young super star clusters 
at D$\sim$500~Mpc for an assumed angular resolution of
0.1$\arcsec$.  For our current best cosmological model, $H_0 =$
65~km~s$^{-1}$~Mpc$^{-1}$, $\Omega_{matter} =$0.3 and $\Lambda =$0.7,
we require an angular resolution of about 10~mas to reach the 0.1~kpc
size scale at moderate redshifts. This would allow us to distinguish
individual luminous star clusters from their surroundings. Finally,
sensitivity becomes a factor at large distances; e.g., an isolated
young globular cluster will have K$\approx$ 29~mag at a redshift of z$\sim$5
(Burgarella, Chapelon, \& Buat 1997).

\begin{table}
\caption{Angular Resolution and Cluster Structures}
\begin{tabular}{lcll}
\tableline
Domain & Radius & 0.1$\arcsec$ Scale  & Results \\
\tableline
Milky Way & 250 kpc & $\leq$ 0.1~pc & low mass stars \\
Local Group & 1 -- 2~Mpc & $\leq$ 1~pc & luminous stars \\
Local Supercluster & $\sim$ 20~Mpc & $\leq$ 10~pc & cluster structures \\
Universe nearby & $\leq$ 500~Mpc & $<$ 0.2~kpc & cluster census \\
High redshift & $z \geq$ 1 & $\sim$ 1~kpc & starburst clumps \\
\tableline
\tableline
\end{tabular}
\end{table}

\section{New Capabilities}

We are benefiting from the continuing technological revolution in astronomy. 
Its components include high sensitivity multi-wavelength coverage; huge 
increases in effective area in the optical and infrared (OIR)
from the construction of 6 to 10-m class telescopes on the ground
and large area observatories in space, such as XMM;
angular resolution gains from {\it HST}, Chandra in X-rays, 
ground-based adaptive optics, 
and radio interferometers; and tremendous strides in numerical modelling 
linked to advances in computing. All 
of these capabilities lead to better measurements of extragalactic 
clusters, as well as improved opportunities to quantitatively 
model their behavior. 

So where do we go from here? Figure 2 illustrates some recent and
planned observational capabilities in terms of wavelength coverage and
angular resolution. For extragalactic star cluster research, where we
need both sensitivity and angular resolution (e.g., super star clusters in 
NGC~1569: de Marchi et al.\ 1997; Hunter et al. 2000; Origlia et al.\
2001; Maoz, Ho, \& Sternberg 2001), the situation looks good
across most of the spectrum. The exception is for improvements in far
ultraviolet studies, where {\it HST} will continue to be the primary
facility. While gains will be achieved in imaging from the Advanced
Camera for Surveys in 2002 and from the Cosmic Origins Spectrograph
(COS) in about 2004, this spectral region, that is critical for
studies of youthful extragalactic star clusters in the nearby
universe, may see little other improvement. Another concern is for future
high angular resolution X-ray capabilities (see Fabbiano 2001).

\begin{figure}
\centerline{\vbox{
  \psfig{figure=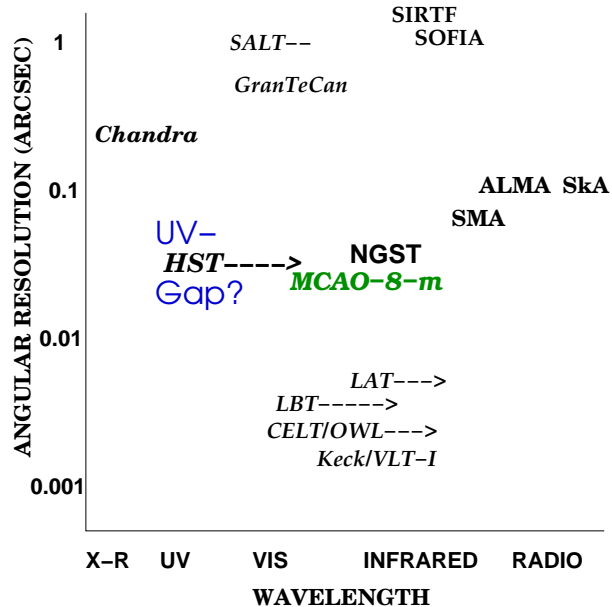,width=8cm,angle=0}
}}
\caption{A summary of some recent and planned observatories, whose 
capabilities can be applied to studies of extragalactic star clusters.
These are discussed in \S3 and \S4 of this paper.}
\end{figure}

\section{New Horizons}

{\bf Formation of Massive Clusters.}  How star clusters, and
especially dense, massive ones, form remains a problem. While progress
has occurred in the development of basic theoretical ideas (e.g.,
Elmegreen \& Efremov 1997; Nakasato, Mori, \& Nomoto 2000; Klessen
2001), observations of super star clusters in formation do not
exist. This issue is further complicated by the lack of even a basic
theoretical model for the formation of high mass stars. Placing
observational constraints on how massive stars (and their associated
star clusters) are made stands as a high observational priority for
ALMA (e.g., Testi 2001), and for sub-mm measurements of the expected
dusty cocoons surrounding the dense molecular cores where the cluster
stars will condense (e.g., Motte \& Andr\'e 2001). Of course, very few, if
any Galactic examples of massive young star clusters exist, and the
work must be carried out in other galaxies. These offer the benefit of
allowing the environments of young clusters to be mapped in
straightforward ways.  The combination of sensitivity and angular
resolution offered by the Sub-Millimeter Array (SMA), now under
construction in Hawaii, then offers substantial observational
opportunities.

{\bf Impacts on the ISM.}  We are seeing progress in observing the
emergence of massive young clusters from their natal gas. Centimeter
wavelength radio continuum observations can detect the optically thick
free-free emission from young, dense HII regions (Turner, Ho, \& Beck
1998, Kobulnicky \& Johnson 1999), and a wider range of radio
techniques apply to the emission from warm molecular gas and likely
presence of masers. With the substantial energy inputs from massive
stars, FIR luminosities are sufficient to allow mapping studies
(Gorjian, Turner, \& Beck 2001).  Youthful FIR-bright super star
clusters could be charted in nearby galaxies with the modest angular
resolutions of SOFIA and SIRTF.  During this phase the molecular
shells are shocked and bathed in FIR radiation, so masing can occur,
opening additional observing strategies for radio interferometers
(Baudry \& Brouillet 1996, Plume et al.\ 1997).  These will benefit
from the increased capabilities of cm wavelength telescope arrays,
such as the enhanced VLA project or future square kilometer array
(SkA).  In the later birth phases, the opacity of the molecular cocoon
drops, and shorter wavelength thermal IR and line emission
(e.g., from shocked H$_2$ or H-Bracket recombination series)
escapes. High angular resolution studies with future giant telescopes,
such as the Large Atacama Telescope (LTA; 15-m class; 35~mas
resolution at K with AO), the 30-m class California Extremely Large
Telescope (CELT), or the even larger ESO 50--100~m OWL telescope hold great
promise.

Once the young cluster is rid of its birth cloud, the ionization
front powered by Lyman continuum radiation rapidly expands, leading to
a giant HII region (or revitalizing one if already present), as in 30
Dor (but some giant HII regions, such as NGC~346 in
the SMC, contain diffuse populations of OB stars). Due to
their large sizes and high monochromatic brightnesses in
H$^+$ recombination and forbidden emission lines, giant HII
regions should be resolvable by a 6.5-m NGST throughout the visible
universe. The bluest bright nebular emission line, [OII]
$\lambda$3727, passes through the inner solar system minimum sky
brightness region near 2~$\mu$m at z$\approx$5, which may represent a
practical, but extremely interesting upper redshift limit for such
studies.

Even before the ionization front breaks out from the molecular gas
around a young star cluster, shocks associated with stellar winds will
develop, and may produce X-rays from embedded star clusters (Hofner \&
Churchwell 1997). Later the colliding stellar winds themselves
(Cant\'o, Raga, \& Rodr\'iguez 2000), as well as any high mass X-ray
binaries (HMXRBs) \footnote{Since many massive stars are in clusters
for much of their lives, we may expect a correlation between HMXRBs
and young super star clusters, although it is not clear if this is
seen.}, will join with supernovae to produce X-rays (Strickland
\& Stevens 1999; Fabbiano, Zezas, \& Murray 2001; Yusef-Zadeh et
al.\ 2001). UV metal resonance line absorption measurements (e.g., with
{\it HST} $+$ COS) offer the possibility to characterize the net
wind from the cluster and its interaction with the surrounding ISM
(Heckmann et al.\ 2001).

{\bf Cluster Structures.} Structures are best derived by
inverting maps of radial distributions of stars and their
3-dimensional velocities calculated from a combination of radial
velocity and proper motion measurements. This process yields
fundamental parameters, such as cluster dynamical masses and
mass-to-light ratios, which tell us about the stellar mass functions
(e.g., Sternberg 1998).  New capabilities have allowed mass
determinations for extragalactic star clusters from measurements of
sizes on {\it HST} images and radial velocity dispersions observed
with sensitive echelle spectrographs on large telescopes (Ho \&
Filippenko 1996; Dubath \& Grillmair 1997; Larsen et al. 2001; Smith
\& Gallagher 2001).

These investigations
can be extended if, for example, we gain the ability to
measure proper motions with high precision in the Milky Way and its
satellites via SIM or GAIA, or can do stellar kinematics with MCAO
\footnote{Multi-conjugate adaptive optics (MCAO) opens the possibility
of producing near diffraction-limited images in the near IR over
moderate fields of view from large ground-based telescopes; see Rigaut
et al. (1999).} feeding efficient spectrographs on very large
telescopes.  An MCAO-equipped 30-m telescope would allow exploration
of the kinematics of individual pre-main sequence stars around young
LMC clusters, permit radial velocity measurements to be routinely made
for evolved stars in clusters within the Local Super Cluster, and
extend routine internal
velocity dispersion measurements to systems of clusters at
D$>$100~Mpc.

Eventually the quest for better angular resolution will
rely on interferometric techniques in the OIR, as well as
throughout the radio.  For example, the Large Binocular Telescope
(LBT) provides 20~mas resolution at K-band from the interferometric
focus of its two 8.4~m mirrors whose centers are separated by
14.4~m (Herbst et al.\ 2002). This resolving power in combination 
with its equivalent of
12-m collecting area is well-matched to half light radii measurements
of compact star clusters throughout the Local Supercluster.  The
ambitious Keck and VLT interferometers hold the promise for order of
magnitude better angular resolution, and when combined with sufficient
sensitivity (a serious technical challenge!) could open a new field of
{\it in situ} studies of young massive clusters at redshifts of
z$\sim$1 (Gallagher \& Tolstoy 1997).

{\bf Stellar Populations.} Massive star clusters are the best representatives 
of "simple stellar populations" and provide fundamental tests of stellar
evolution models and their offspring, the stellar population synthesis
codes (see D'Antona 2002). These checks rest on our ability to
measure color-magnitude diagrams, which are primarily a matter of
combining sufficient resolution with photometric accuracy. This is
difficult in cluster cores in the nearest galaxy, the LMC, 
even with {\it HST}; so we still have a ways to go.  However, when 
multiple colors covering a wide wavelength baseline can be obtained, 
the statistics of cluster ages can be determined from their integrated 
broad-band colors, for example, by de Grijs, O'Connell, \& Gallagher (2001)
in M82.

A second technique relies on comparisons of synthetic and observed spectral
energy distributions (e.g., Gonz\'alez-Delgado, Leitherer, 
\& Heckman 1999). This approach is ripe for exploitation with the
advent of large format detectors coupled to multi-object and echelle
spectrographs on high performance large telescopes. 
It will further benefit from the growth in numbers of 10-m aperture 
optical telescopes, such as the GranTeCan project, as well as the 
spectroscopically oriented Hobby-Eberly Telescope (HET), and its more 
UV-sensitive off-spring, the Southern African Large Telescope 
(SALT), now under construction. Current studies have
focused on determinations of ages and stellar luminosity (and thus
mass) functions from optical spectra (e.g., Gallagher \& Smith 1999,). 

Since AO works best in the NIR, a new emphasis is being placed on the
analysis of H- and K-band cluster spectra (e.g., Lan\c con et
al.\ 1999; Alonso-Herrero, Ryden \& Knapen 2000; Mengel et al.\ 2001).
Ages of star clusters can be estimated from stellar absorption
features in IR spectra especially by including 
the impact of the extended-AGB intermediate age stellar
populations (see Lan\c con \& Mouchine 2001). Still in its infancy but
astrophysically critical is to derive ages {\it and} abundances from
the integrated light of younger star clusters (Maraston et
al.\ 2001). This exciting step would extend ``Galactic'' approaches for
investigating the evolution of stellar populations to a variety of
galaxies.

{\bf Disruption.} We close our glimpse of the future by considering
a nagging question: are massive young super star
clusters analogous to common globular star clusters in their youth?
The most direct way to resolve this issue would be to characterize youthful
globular star clusters by observing them at high redshifts.
Unfortunately, this is extremely difficult to do. 

Conversely, we can ask if the massive young clusters we see today are
likely to survive to ages of $>$8 -- 12~Gyr?  Then having resolved this 
basic point, we
could turn to the more general problem emphasized by, among others,
Fritze v.-Alvensleben (1998), Whitmore et al. (1999), and Vesperini
(2001): whether {\it systems} of young super star clusters would
evolve to have the characteristics (e.g., luminosity function) of the
populations of ancient globular clusters now seen in galaxies.

A difficulty with this approach is that the present-day globular
cluster systems probably represent small fractions of initially much
larger numbers of clusters, with many clusters having been lost
through various destruction mechanisms (Gnedin \& Ostriker 1997,
Vesperini 2000).  We are beginning to benefit from wide angle
observations, which give quantitative insights into ongoing 
destruction processes (Odenkirchen et al.\ 2001; Siegel et
al.\ 2001).  More information will become available on Milky Way
clusters from, for example, the Sloan Digital Sky Survey (SDSS), and
with the added attraction of the Magellanic Clouds from the southern
hemisphere Visible and Infrared Survey Telescope (VISTA), as well as
from the NIR Primordial Explorer (PRIME) in space.

Further progress on the cluster survival issue comes from combining
advances in theory and observations.  Improved numerical models yield
better information on the predicted behavior of massive clusters as
they respond to internal effects, such as mass loss driven by stellar
evolution and the formation and destruction of binary stars, as well
as externally imposed dynamical constraints (Aarseth 1999; Kim,
Morris, \& Lee 1999; Fukushige \& Heggie 2000; Giersz 2001 and
references therein). A recent application of this approach to star
clusters located near the center of the Milky Way shows that these
objects cannot survive for long ($\ll$1~Gyr) unless they are very
massive (Kim et al.\ 2000; Portegies Zwart et al.\ 2001). Additional
boundary conditions are imposed by the stellar mass functions; clusters with
flatter mass functions are less likely to survive (e.g., Takahashi \&
Portegies Zwart 2000 and references therein). This could be an
important factor since some star clusters in dense environments show
signs of relatively flat stellar mass functions (Figer et al.\ 1999;
Smith \& Gallagher 2001), complicating the use of cluster luminosities
to trace masses and thus destruction probabilities. Studies of
intermediate age merger remnants, such as NGC~1316 by Goudfrooij et al.\
(2001), provide a critical link between the recent starbursts and the
globulars, and give reassurance that some massive star clusters born
in starbursts survive well into middle age.

It is definitely premature to consider this or many other fundamental
issues regarding star clusters in our own and other galaxies as
a closed subject.  Rather this is an area where we may expect
substantial progress in the coming years. 

\bigskip

JSG thanks NASA for their funding of cluster research through 
programs associated with the {\it Hubble Space Telescope}, and the 
University of Wisconsin-Madison Graduate School for support from 
the Vilas trust.  We thank all the sponsors of this
IAU Symposium for their generous support, including the large 
observatories in Chile (AUI/NRAO, CTIO, ESO, Gemini, and LCO),
whose current and future instrumentation will make possible much of the 
research described above.

\end{document}